\documentclass[%
superscriptaddress,
 amsmath,amssymb,
 aps,twocolumn,
floatfix,
]{revtex4-2}
\usepackage{colortbl}
\usepackage{color}
\usepackage{amsfonts,amsmath,amssymb}
\usepackage[dvipsnames]{xcolor}
\usepackage{tikz}
\usepackage{tabularx,hhline}
\usepackage{bm}
\usepackage{placeins}
\usepackage{dcolumn,stmaryrd}
\usepackage[version=4]{mhchem}
\usepackage[colorlinks=True,linkcolor=blue,citecolor=blue,urlcolor=blue]{hyperref}

\usetikzlibrary{calc}
\newcommand{\zn}{ZnCr$_2$O$_4$}
\newcommand{\mg}{MgCr$_2$O$_4$}

\begin{document}
\title{Spin-Peierls Transition in the frustrated spinels ZnCr$_{2}$O$_{4}$ and MgCr$_{2}$O$_{4}$}

\author{Ludovic D. C. Jaubert}
\affiliation{CNRS, University of Bordeaux, LOMA, UMR 5798, F-33400 Talence, France}
\affiliation{Department of Physics and Quantum Centre of Excellence for Diamond and Emergent Materials (QuCenDiEM), Indian Institute of Technology Madras, Chennai 600036, India}

\author{Yasir Iqbal}
\affiliation{Department of Physics and Quantum Centre of Excellence for Diamond and Emergent Materials (QuCenDiEM), Indian Institute of Technology Madras, Chennai 600036, India}

\author{Harald O. Jeschke}
\affiliation{Research Institute for Interdisciplinary Science, Okayama University, Okayama 700-8530, Japan}
\affiliation{Department of Physics and Quantum Centre of Excellence for Diamond and Emergent Materials (QuCenDiEM), Indian Institute of Technology Madras, Chennai 600036, India}

\begin{abstract}
The chromium spinels \ce{MgCr2O4} and \ce{ZnCr2O4} are prime examples of the highly frustrated pyrochlore lattice antiferromagnet. Experiment has carefully established that both materials, upon cooling, distort to lower symmetry and order magnetically. We study the nature of this process by a combination of density-functional-theory based energy mapping and classical Monte Carlo simulations. We first computationally establish precise Heisenberg Hamiltonian parameters for the high temperature cubic and the low temperature tetragonal and orthorhombic structures of both spinels. We then investigate the respective ordering temperatures of high symmetry and low symmetry structures. We carefully compare our results with experimental facts and find that our simulations are remarkably consistent with a type of spin-Peierls mechanism, adapted to three dimensions, where the structural distortion is mediated by a magnetic energy gain due to a lower degree of frustration.
\end{abstract}

\date{\today}
\maketitle
\begin{figure}[ht]
\centering\includegraphics[width=0.95\columnwidth]{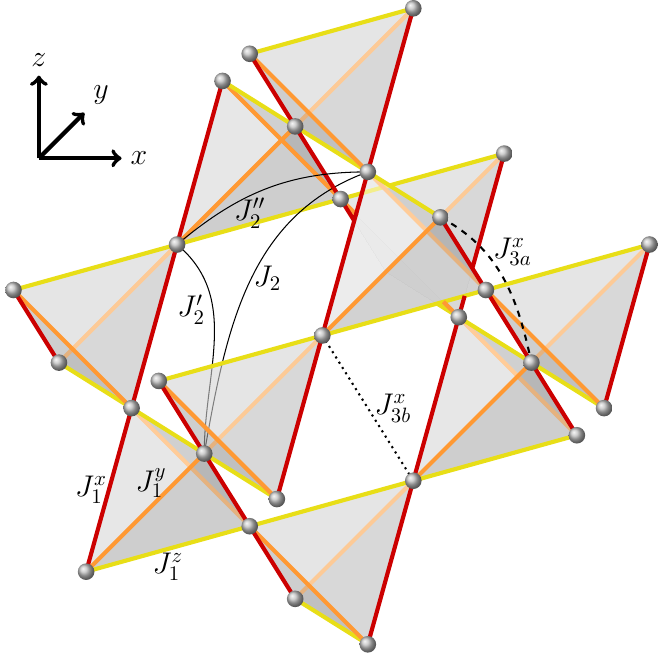}
\caption{
Illustration of the inequivalent exchange couplings for the orthorhombic space group of the pyrochlore lattice. The two types of third neighbor bonds $J_{3a,b}^\alpha$ are orthogonal to the $\alpha\in\{x,y,z\}$ cubic axis. In the tetragonal structure, $J^x=J^y$ between first and third neighbors, and $J_2'=J_2''$. In the cubic structure, the $x,y$ and $z$ axes are equivalent.
}
\label{fig:paths}
\end{figure}

{\it Introduction --} The physics of highly frustrated magnets has been a focus of condensed matter theory and experiment for about half a century. The structure with arguably the largest diversity of materials is the pyrochlore lattice [see Fig.~\ref{fig:paths}]. Chromium spinels, and in particular {\zn} and {\mg}, have long been studied as some of the best realizations of the nearest neighbor antiferromagnet on the pyrochlore lattice. Among the oxides, {\zn} and {\mg} have the smallest lattice parameters and thus the smallest Cr-Cr separations~\cite{Tsurkan2021}. This leads to large antiferromagnetic nearest neighbor interactions as demonstrated by the large negative Curie-Weiss temperatures of $\theta_{\rm CW}=-390$\,K and $\theta_{\rm CW}=-400$\,K, respectively~\cite{Takagi2011}. While the classical Heisenberg antiferromagnet on the pyrochlore lattice is theoretically predicted to realize a classical spin liquid~\cite{Moessner1998}, {\zn} and {\mg} order magnetically around the same temperature $T_{\rm N}\approx 12-13$\,K~\cite{Lee2000,Rovers2002}. The strong magnetic frustration can be recognized from the fact that the ratio between Curie-Weiss and N{\'e}el temperatures $f=|\theta_{\rm CW}|/T_{\rm N}$ is above 30 for both materials. Also, the full entropy of the spin system $S_m=2R\ln(2S+1)$ is only reached above room temperature~\cite{Kant2009}. While the presence of significant spin-lattice coupling and a structural transition at $T_{\rm N}$ was realized early on~\cite{Kino1971}, the low temperature structures were identified only recently~\cite{Ortega2008,Tomiyasu2008,Kemei2013,Tomiyasu2013,Gao2018,Bai2019}. The concomitant magnetic ordering and structural phase transitions have been interpreted as a spin-Peierls transition~\cite{Lee2000,Tchernyshyov2002,Fennie2006,Xiang2011,Aoyama-2016}.

\begin{figure*}[ht]
\centering
\includegraphics[width=0.95\textwidth]{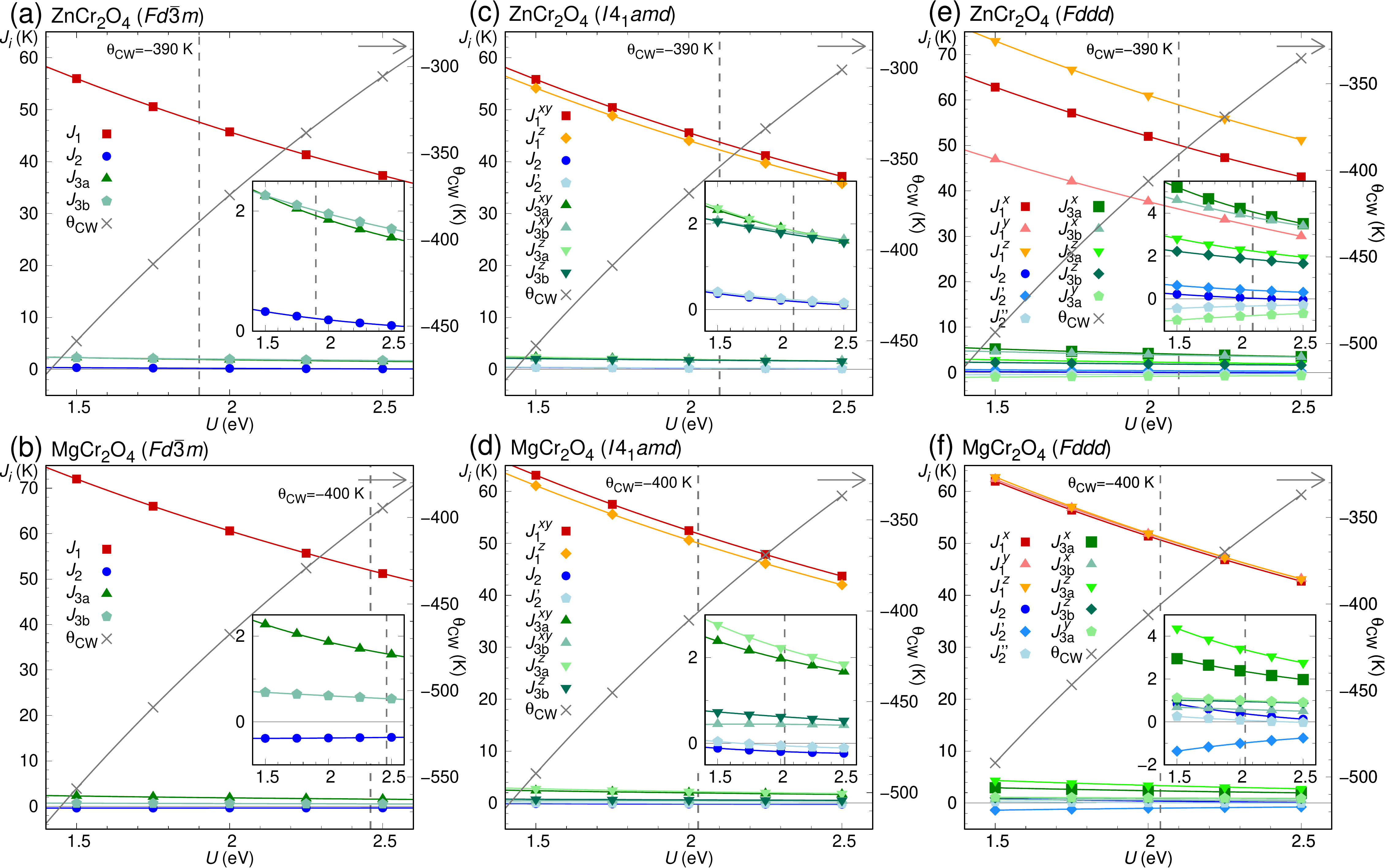}
\caption{Exchange couplings of {\zn} and {\mg} extracted with GGA+U at $J_H=0.72$~eV~\cite{Mizokawa1996} as function of interaction strength $U$. (a) and (b) for cubic, (c) and (d) for tetragonal and (e) and (f) orthorhombic structure. The Curie-Weiss temperatures are shown as grey lines (see Eqs.\ (\ref{eq:cubic},\ref{eq:tetra},\ref{eq:ortho})), and vertical dashed lines indicate the interpolated Heisenberg Hamiltonian parameters which match the experimental value of $\theta_{\rm CW}$.}
\label{fig:couplings}
\end{figure*}

In this letter, we develop an original approach to understand this transition straddling frustrated magnetism and magneto-elastic distortion, using a combination of Density Functional Theory (DFT) and Monte Carlo simulations, based on experimental data. First, we compute the ab initio spin exchange Hamiltonian of these two materials for the three distinct crystal structures that have been found in experiments: cubic ($Fd\bar{3}m$), tetragonal ($I4_1/amd$) and orthorhombic ($Fddd$). {\mg} and {\zn} are known to be close to the pyrochlore antiferromagnetic spin liquid, whose extensive entropy strongly hinders any ordering mechanism. Quantitative results around the transition thus necessarily require an approach beyond mean field: We simulate these ab initio Hamiltonians with a Monte Carlo algorithm, able to properly account for the thermal fluctuations of this strongly correlated magnetic texture. Our main results are that, despite an exchange energy scale of $\theta_{\rm CW}\sim 400$ K, and distinct Hamiltonians derived from independent experimental data, our simulations remarkably fit the experimental transition temperature for \textit{both} materials within a few Kelvin only. The models with cubic (resp. tetragonal and orthorhombic) symmetry have a transition temperature below (resp. above) the one found in experiments, $T_{\rm N}\approx 12.5\pm 0.5$ K. In addition, we confirm that the low-symmetry systems are magnetically more stable at $T_{\rm N}$. Our results are consistent with a spin-Peierls scenario where the cubic symmetry persists down to $T_{\rm N}$, where the structural transition into a lattice with lower symmetry is induced by a gain in magnetic exchange energy and accompanied by long-range magnetic order at the same temperature $T_{\rm N}$.\\

\begin{table*}[htb]
\begin{tabular}{c|c|c|c|c|c|c|c|c|c|c|c|c}
\multicolumn{13}{c}{cubic $Fd\bar{3}m$} \\
material& $J_1$\,(K)&&&    $J_2$\,(K)&&&     $J_{3a}$\,(K)&   $J_{3b}$\,(K) &&&\\\hline
{\zn}&{47.6(3)}&&&{0.2(2)}&&&{1.9(2)}&{2.0(2)}&&&\\
{\mg}&{51.9(2)}&&&{-0.4(1)}&&&{1.6(2)}&{0.5(2)}&&&\\
\multicolumn{13}{c}{tetragonal $I4_1/amd$} \\
material& $J_1^{xy}$\,(K)&    $J_1^z$\,(K)&&   $J_2$\,(K)&   $J_2'$\,(K)&&   $J_{3a}^{xy}$\,(K)&   $J_{3b}^{xy}$\,(K) &$J_{3a}^z$\,(K) &$J_{3b}^z$\,(K) &&\\\hline
{\zn}&{48.2(2)}&{46.6(2)}&&{0.3(4)}&{0.3(1)}&&{2.0(3)}&{1.9(3)}&{2.0(3)}&{1.9(3)}&& \\
{\mg}&{51.8(2)}&{50.0(2)}&&{-0.2(4)}&{-0.1(1)}&&{2.0(3)}&{0.4(3)}&{2.2(3)}&{0.6(3)}&&\\
\multicolumn{13}{c}{orthorhombic $Fddd$} \\
material& $J_1^x$\,(K)&    $J_1^y$\,(K)&   $J_1^z$\,(K)&   $J_2$\,(K)&  $J_2'$\,(K)&  $J_2''$\,(K)&   $J_{3a}^x$\,(K)&   $J_{3b}^x$\,(K) &$J_{3a}^z$\,(K) &$J_{3b}^z$\,(K) & $J_{3a}^y$\,(K)& $J_{3b}^y$\,(K)\\\hline
{\zn}&{49.9(2)}&{35.9(2)}&{58.8(2)}&{0.0(5)}&{0.4(2)}&{-0.3(5)}&{4.1(2)}&{3.8(2)}&{2.2(2)}&{1.9(2)}&{-0.8(2)}& --\\
{\mg}&{50.6(2)}&{51.0(3)}&{51.2(3)}&{0.4(9)}&{-1.0(9)}&{0.1(3)}&{2.4(3)}&{0.6(2)}&{3.4(5)}&{0.9(4)}&{1.0(3)}& -- \\
\end{tabular}
\caption{
Exchange couplings of cubic, tetragonal and orthorhombic {\zn} and {\mg}, calculated within GGA+U. These sets of couplings correspond to the vertical lines in Fig.~\ref{fig:couplings} and match the experimental Curie-Weiss temperatures $\theta_{\rm CW}=-390$\,K for {\zn}~\cite{Martinho2001,Takagi2011} and $\theta_{\rm CW}=-400$\,K for {\mg}~\cite{Rovers2002,Takagi2011}, calculated using Eqs.~\eqref{eq:cubic}-\eqref{eq:ortho}.
}
\label{tab:couplings}
\end{table*}

{\it Determination of the ab-initio Hamiltonians --} We study the following Hamiltonian
\begin{equation}
    H=\sum_{i<j} J_{ij} {\bf S}_i\cdot {\bf S}_j \,,
\label{eq:ham}
\end{equation}
where the sum runs over first-, second- and third-neighbour pairs of spins on the pyrochlore lattice. The large spins $S=3/2$ of the Cr$^{3+}$ magnetic ions are approximated as classical Heisenberg spins. In order to determine the values of the $J_{ij}$ couplings, we perform electronic structure calculations for {\mg} and {\zn} using the following structures: Above the N\'eel temperature, {\zn} and {\mg} are known to crystallize in the cubic $Fd\bar{3}m$ (No. 227) structure. We use lattice constants measured at $T=15$~K for both materials from Ref.~\onlinecite{Dutton2011}. Below $T_{\rm N}$, it remains unclear whether their structures deform into tetragonal $I4_1/amd$ (No. 141), orthorhombic $Fddd$ (No. 70), or a mixture of both. For the sake of completeness, we perform DFT calculations for both structures. We use the tetragonal structures of {\zn} measured with powder X-ray diffraction at $T=5.4$~K  from Ref.~\onlinecite{Kemei2013}, and of {\mg} measured with neutron diffraction at $T=10$~K from Ref.~\onlinecite{Ortega2008}. Finally, we use the orthorhombic $Fddd$ (No. 70) structure of {\zn} and {\mg} measured with powder X-ray diffraction at $T=5.7$~K and $T=5.4$~K, respectively, from Ref.~\onlinecite{Kemei2013}.

The exchange couplings for cubic, tetragonal and orthorhombic space groups of {\zn} and {\mg} are illustrated in Fig.~\ref{fig:paths}. The symmetry lowering from cubic to tetragonal structure leads to the six nearest neighbor bonds in the Cr tetrahedron splitting into four shorter and two slightly longer bonds, here called $J_1^{xy}$ and $J_1^{z}$. In our naming scheme, $J_1^{z}$ indicates that these bonds are perpendicular to the $z$ direction, and $J_1^{xy}$ bonds are all perpendicular to either $x$ or $y$ directions. The twelve next-nearest neigbors of the cubic structure $J_2$ split into four $J_2$ and eight $J_2'$ bonds. And as usual on the pyrochlore lattice, two types of third-neighbor exchange paths exist: either the six $J_{3a}$ bonds with a Cr site in between or the six $J_{3b}$ bonds across hexagons. Each $J_{3a,b}$ bond belongs to a plane orthogonal to a cubic axis and can be labeled $J_{3a,b}^{xy}$, $J_{3a,b}^{z}$. Symmetry lowering to the orthorhombic structure leads to three inequivalent exchange paths for nearest neighbors, three second-neighbor paths and six third-neighbor paths, differentiating the $x$ and $y$ axes.

\begin{figure*}[ht]
\centering
\includegraphics[height=8.5cm]{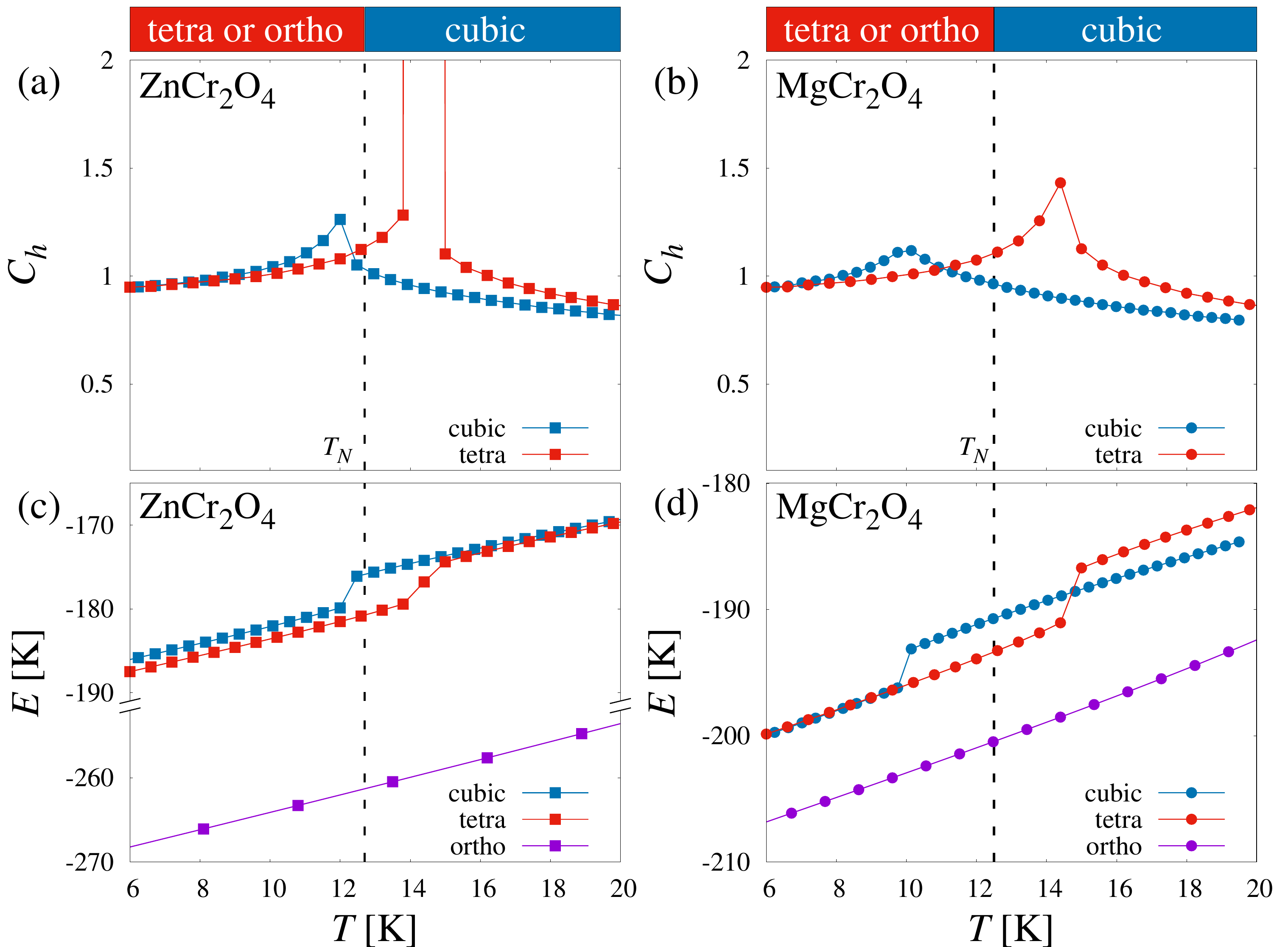}
\includegraphics[height=8.5cm]{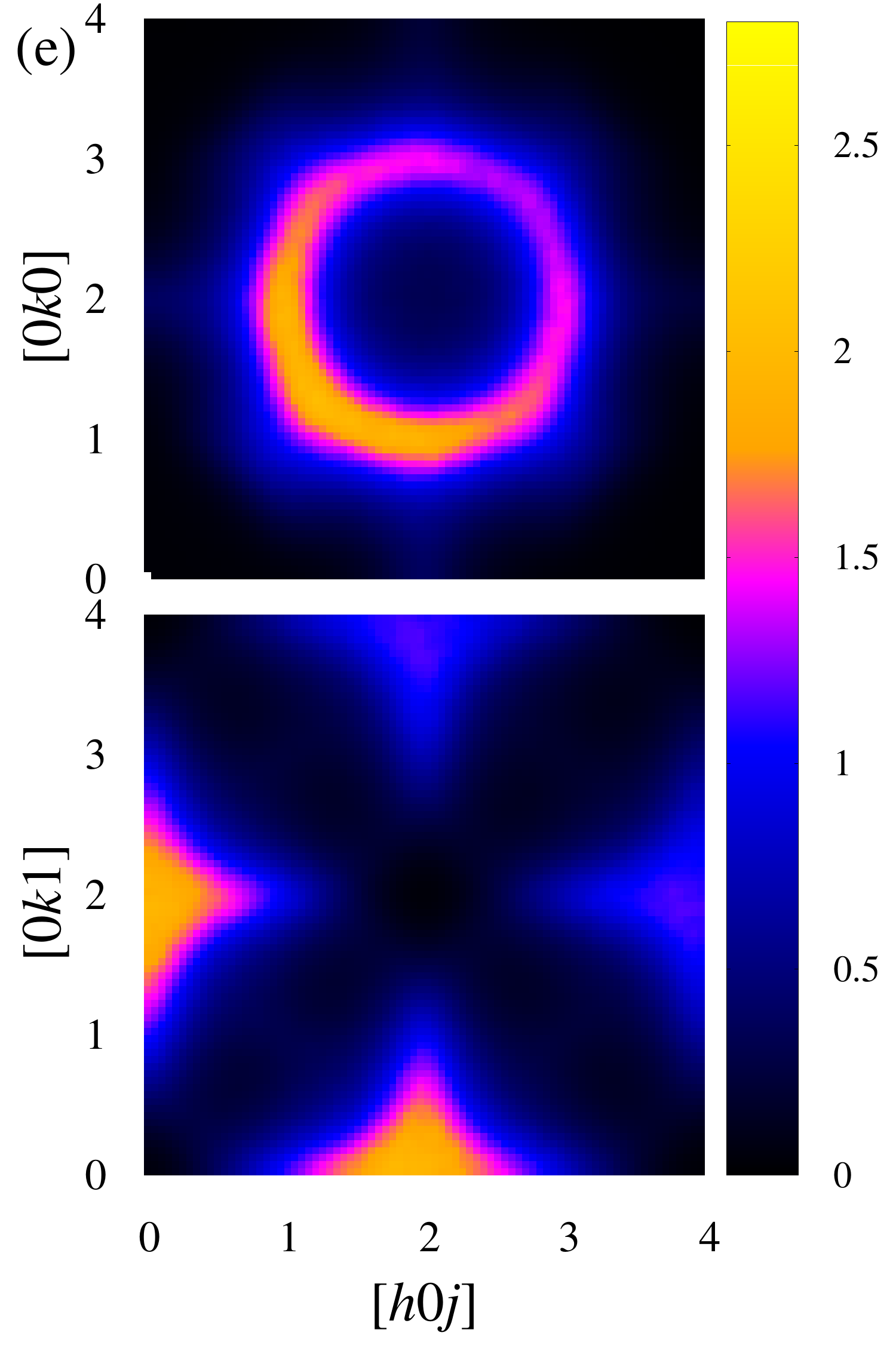}
\caption{
\textbf{Monte Carlo simulations:}
Specific heat (a,b) and energy per spin (c,d) for parameters of {\zn} (a,c) and {\mg} (b,d) given in Table \ref{tab:couplings} and for cubic (blue), tetragonal (red) and orthorhombic (violet) lattice symmetry. The dashed line is the position of the N\'eel temperature $T_N$ measured experimentally in each material \cite{Lee2000,Rovers2002}. $C_h$ for the orthorhombic structure is not plotted because its transition temperature is higher than 20K: 70 K and 25 K for {\zn} and {\mg} respectively.
(e) Structure factor as seen by neutrons for MgCr$_{2}$O$_{4}$ at 20 K in the structurally cubic phase (\textit{Fd$\bar{3}$m}) in the [$hkj$] plane for $j = 0$ (top) and $1$ (bottom).
}
\label{fig:MC}
\end{figure*}

We perform DFT energy mapping for the six crystal structures in order to determine the parameters of the Heisenberg Hamiltonian (\ref{eq:ham}). This approach has already been successful for other Cr$^{3+}$ materials with $S=3/2$~\cite{Tapp2017,Ghosh2019,Gen2023}. All electronic structure calculations are based on the full potential local orbital (FPLO) method~\cite{Koepernik1999} in combination with a generalized gradient approximation (GGA) exchange correlation functional~\cite{Perdew1996}. We use collinear DFT here; effects of spin-orbit coupling like the Dzyaloshinskii-Moriya (DM) interaction that would require non-collinear DFT can be expected to be small in Cr$^{3+}$ due to the half-filled $t_{2g}$ shell, and a calculation for {\mg} has found DM to be negligible~\cite{Xiang2011}{, while previous theoretical fits to neutron-scattering measurements did not require any anisotropic terms \cite{Lee2002,Bai2019}. At lower energy scale, anisotropy could possibly play a role to understand the subtle degeneracy lifting \cite{Bai2019,Glittum2023} responsible for the magnetic textures at very low temperatures. 
While this question is beyond the scope of our present Letter, non-collinear DFT methods \cite{Szilva2013} could be useful in that context}. We determine the exchange interactions by calculating DFT energies for a number of spin configurations that is substantially larger than the number of couplings we want to determine. This requires a $2\times 1\times 1$ supercell with eight symmetry inequivalent Cr moments in the case of the cubic structure and  $2\times 2\times 1$ supercells with sixteen symmetry inequivalent Cr moments in the case of tetragonal and orthorhombic structures. The results are shown graphically in Fig.~\ref{fig:couplings}. In order to determine the experimentally appropriate value of the $U$ parameter for Cr $3d$ in the DFT+U calculations~\cite{Liechtenstein1995}, we demand that the Hamiltonian couplings match the experimentally determined Curie-Weiss temperatures $\theta_{\rm CW}=-390$\,K for {\zn}~\cite{Martinho2001,Takagi2011} and $\theta_{\rm CW}=-400$\,K for {\mg}~\cite{Rovers2002,Takagi2011}. The  Curie-Weiss temperatures are estimated for the cubic structure as
\begin{equation}
\theta_{\rm CW}=-\frac{1}{3} S (S + 1) \big(6 J_1 + 12 J_2 + 6 J_{3a} + 6 J_{3b}\big)\,,
\label{eq:cubic}
\end{equation}
for the tetragonal structure as
\begin{eqnarray}
\begin{split}
\theta_{\rm CW}=-\frac{1}{3} &S (S + 1) \big(4 J_1^{xy} + 2J_1^z + 4 J_2 + 8 J_2' \\&+ 4 J_{3a}^{xy} +   4 J_{3b}^{xy} + 2 J_{3a}^z + 2 J_{3b}^z + 4 J_7\big)\, ,
\label{eq:tetra}
\end{split}
\end{eqnarray}
and for the orthorhombic structure as
\begin{equation}\begin{split}
\theta_{\rm CW}=-\frac{1}{3} &S (S + 1) \big(2J_1^x +2J_1^y + 2J_1^z + 4 J_2 + 4 J_2' + 4 J_2'' \\&+ 2 J_{3a}^x + 2J_{3b}^x +2 J_{3a}^z +2 J_{3b}^z +2 J_{3a}^y\big)
\label{eq:ortho}
\end{split}\end{equation}
The selected $U$ values are between 2 and 2.5\,eV and comparable with values found to be relevant for other chromium spinels~\cite{Ghosh2019}. There is a certain variation in the reported Curie-Weiss temperatures of {\zn}~\cite{Plumier1977,Ueda2007,Martinho2001,Rudolf2007} and {\mg}, but as the evolution of exchange interactions with $U$ is rather smooth and ratios do not vary much, the results are not very sensitive to the choice. For the cubic structure, comparable calculations have been performed before~\cite{Yaresko2008,Wysocki2016,Xiang2011}. However, for {\zn}, Ref.~\onlinecite{Yaresko2008} does not separate $J_{3a}$ and $J_{3b}$, and Ref.~\onlinecite{Wysocki2016} reports identical Hamiltonians for cubic {\zn} and {\mg}. Our results for cubic {\mg} are consistent with ab-initio calculations of Ref.~\cite{Xiang2011} and can be compared to the determination of the Heisenberg Hamiltonian parameters by Bai {\it et al}.~\cite{Bai2019} using fits to inelastic neutron scattering data; they find the set of couplings $\{J_1, J_2, J_{3a}, J_{3b}\}$ to be  $\{38.05(3),3.10,4.00,0.32\}$\,K. We do not expect to find a perfect match with ab-initio calculations, but it is particularly encouraging for our DFT results to recover the same main physics: We also find a dominant antiferromagnetic $J_1$ coupling, one order of magnitude larger than the second exchange path $J_{3a}$ (also antiferromagnetic) and with a small $J_{3b}$ term. There is a difference though, in that the $J_2$ term changes sign. This exchange path is, however, known to be subdominant, with previous studies disagreeing on its sign \cite{Wysocki2016,Xiang2011,Bai2019}. For the sake of completeness, we have checked that using Bai's parameters does not affect the conclusions of our paper \cite{supp}. In addition the structure factor obtained by Monte Carlo simulations at 20\,K [see discussion below and Fig.~\ref{fig:MC}(e)] is consistent with the one measured experimentally in \cite{Bai2019}. We are thus confident that our ab-initio results are set on a sensible experimental footing. To the best of our knowledge, no ab-initio estimates of the tetragonal and orthorhombic Hamiltonian couplings have been published for {\zn} and {\mg}. We chose to use the same value of $\theta_{\rm CW}$ for the three lattice structures of each material as it allows us to impose a constant energy scale of the total magnetic exchange path across the transition. Our results are thus a direct consequence of the lower degree of magnetic frustration in lower-symmetry structures, and cannot be due to an overall rescaling of the exchange couplings.\\

{\it Monte Carlo simulations --} Using the parameters of Table \ref{tab:couplings}, we performed Monte Carlo simulations for {\zn} and {\mg} in their three distinct lattice structures. The proximity of spiral phases, a degenerate manifold with soft-mode excitations and magnetic incommensurability to the model \cite{Bai2019,Glittum2023} renders the determination of the low-temperature magnetic order particularly non-trivial, and it is not the goal of this letter to address this point. Instead we focus on the transition at finite temperature, and how our results, summarised in Fig.~\ref{fig:MC}, fit with experiments.

The transition temperatures for cubic simulations are \textit{lower} than the one found in experiments $T_{\rm N}=12.5\pm 0.5$ K. Simulations thus expect {\zn} and {\mg} to remain magnetically disordered down to $T_{\rm N}$, as observed in experiments. On the other hand, the transition temperatures for tetragonal and orthorhombic simulations are \textit{higher} than $T_{\rm N}$. It means that a structural transition in this temperature range is automatically accompanied by magnetic ordering, as observed in experiments. And at $T_{\rm N}$, the magnetic energy of the system is always lower for the tetragonal and orthorhombic structures than for the cubic one. 

In frustrated magnets, the value of $T_{\rm N}$ sensitively depends on the ratios of the exchange interactions, and is largely independent from $\theta_{\rm CW}$; this is especially true for {\zn} and {\mg} which are very close to the Heisenberg antiferromagnetic spin liquid. In that context, and keeping in mind that we are simulating systems whose exchange energy scales are of the order of $|\theta_{CW}|\approx 400$ K, and whose coupling parameters have been obtained from unbiased DFT calculations derived from independent experimental data (X-ray diffraction and magnetic susceptibility measurements) for two different materials, it is truly remarkable that the transition temperatures of the cubic and tetragonal structures occur within a very small temperature window $\Delta T \sim 5 {\rm K} \ll |\theta_{CW}|$, with the experimental transition temperature $T_{\rm N}$ sitting precisely between them. Our simulations show that both {\zn} and {\mg} remain magnetically disordered down to the necessary temperature for the tetragonal distortion to be able to bring a net magnetic energy gain [Fig.~\ref{fig:MC}(c,d)]. This is consistent with a three-dimensional analog of the spin-Peierls mechanism, where the tetragonal distortion is made possible thanks to the energy gain in magnetic interactions. To this day, the literature remains unclear whether both tetragonal and orthorhombic structures really co-exist, and in which ratio and temperature window. But since the simulated transition temperatures of the orthorhombic structures, respectively 70 K and 25 K for {\zn} and {\mg}, are noticeably higher than in experiments, our simulations suggest that the spin-Peierls mechanism is attached to the tetragonal distortion. In the future, it would be interesting to see if the tetragonal structure could serve as an intermediate state facilitating the onset of an orthorhombic distortion.

Another notable outcome is that the cubic magnetic transition is only a few Kelvin below $T_{\rm N}$, i.e., only a few Kelvin below the experimentally accessible regime with cubic structure. It is thus not unreasonable for Ref.~\cite{Nassar2024} to have recently observed an onset of magnetic order above $T_N$ in {\mg}.

{\it Conclusions}\textendash In this work, we elucidate the subtle interplay between structural distortion and magnetic order in two celebrated pyrochlore antiferromagnets {\zn} and {\mg}, thus answering the riddles surrounding their transitions upon cooling. We identify a spin-Peierls mechanism at play which is responsible for the observed structural transition from the high temperature cubic to the low temperature tetragonal phase. Employing ab-initio density functional theory and classical Monte Carlo simulations, we show that the experimental $T_{\rm N}$ lies precisely within a remarkably tight window which is bounded above and below by the $T_{\rm N}$ of the tetragonal and cubic structures, respectively. It is this very fact which allows for a magnetic energy gain via a tetragonal distortion thus inducing a structural transition and explains the concomitant onset of magnetic order. By dissecting this intertwined nature of structural and magnetic transitions, our work resolves a long-standing enigma of these chromium spinels, using an approach that can easily be applied to other materials with magneto-structural distortion.

{\it Acknowledgements}\textendash The work of Y.I. was performed in part at the Aspen Center for Physics, which is supported by National Science Foundation Grant No. PHY-2210452. The participation of Y.I. at the Aspen Center for Physics was supported by the Simons Foundation (1161654, Troyer). The research of Y.I. was carried out, in part, at the Kavli Institute for Theoretical Physics in Santa Barbara during the “A New Spin on Quantum Magnets” program in summer 2023 and ``Correlated Gapless Quantum Matter" program in spring 2024, supported by the National Science Foundation under Grant No. NSF PHY-2309135. Y.I. acknowledges support from the ICTP through the Associates Programme, from the Simons Foundation through Grant No. 284558FY19, and IIT Madras through the Institute of Eminence (IoE) program for establishing QuCenDiEM (Project No. SP22231244CPETWOQCDHOC). Y.I. also acknowledges the use of the computing resources at HPCE, IIT Madras. L.J. and H.O.J. thank IIT Madras for a Visiting Faculty Fellow position under the IoE program during which this work was initiated. L.J. acknowledges financial support from Grants No. ANR-18-CE30-0011-01 and ANR-23-CE30-0038-01. Y.I. and H.O.J. acknowledge the hospitality of the University of
Bordeaux/CNRS during a work visit in September 2023. All authors thank the International Centre for Theoretical Sciences (ICTS), Bengaluru, India for hospitality during the program ``Frustrated Metals and Insulators'' (Code No. ICTS/frumi2022/9).

\bibliography{crspinels.bib}
\end{document}